\documentclass[10pt]{iopart}

\usepackage{graphicx}

\usepackage{url}

\begin{document}

\title{Magnetic ground state of supported monatomic Fe chains from first principles}

\author{B Nagyfalusi$^{1,2}$, L Udvardi$^{2,3}$ and L Szunyogh$^{2,3}$}

\address{$^1$ Wigner Research Centre for Physics, Institute for Solid State Physics and Optics, H-1525 Budapest, Hungary}
\address{$^2$ Department of Theoretical Physics, Institute of Physics, Budapest University of Technology and Economics, M\H uegyetem rkp. 3., H-1111 Budapest, Hungary}
\address{$^3$ MTA-BME Condensed Matter Research Group, Budapest University of Technology and Economics, M\H uegyetem rkp. 3., H-1111 Budapest, Hungary}
\ead{nagyfalusi.balazs@wigner.hu}

\begin{abstract}
A new computational scheme is presented based on a combination of the conjugate gradient and the Newton-Raphson method to self-consistently minimize the energy within local spin-density functional theory, thus to identify the ground state magnetic order of a finite cluster of atoms. 
The applicability of the new \textit{ab initio} optimization method is  demonstrated for Fe chains deposited on different metallic substrates.
The optimized magnetic ground states of the Fe chains on Rh(111) are analyzed in details and a good comparison is found with those obtained from an extended Heisenberg model containing first principles based interaction parameters. Moreover, the effect of the different bilinear spin-spin interactions in the formation of the magnetic ground states is monitored. In case of Fe chains on Nb(110) spin-spiral configurations with opposite rotational sense are found as compared to previous spin-model results which hints on the importance of higher order chiral interactions. The wavelength of the spin-spiral states of Fe chains on Re(0001) was obtained in good agreement with scanning tunneling microscopy experiments.
\end{abstract}

\noindent{\it Keywords\/}: magnetic nanoclusters, \textit{ab initio}, magnetic ground state

\submitto{\JPCM}
\maketitle

\ioptwocol

\section{Introduction}

The rapid development of scanning tunneling microscopy (STM) allows to create finite clusters of magnetic atoms on solid surfaces with atomic precision \cite{Eigler1990,Meier2008}.
Magnetic chains on top of a superconducting surface has attracted considerable interest during the past years due to the challenge of finding Majorana bound states that may open a new prospect in quantum computation \cite{Andolina2017,Kim2018,beck2021,crawford2021,kuester2021}. 
The magnetic structure of the chains has crucial role in the emergence of topological superconductivity.
A recent theoretical and experimental review on the physics of atomic spin chains deposited on surfaces is provided in \cite{Choi2019}.

Complex magnetic systems are often described in terms of a generalized Heisenberg model, where the interaction between spins $\mathbf{s}_i$ and $\mathbf{s}_j$ at sites $i$ and $j$ is ascribed to a 
tensorial exchange coupling $\mathbf{J}_{ij}$. 
This exchange coupling can further be split into three terms \cite{Udvardi2003}:
the isotropic exchange,  the symmetric anisotropic exchange and the antisymmetric anisotropic exchange or  the Dzyaloshinky-Moriya interaction (DMI) \cite{Dzyaloshinsky1958, Moriya1960}.
Of particular interest is the DMI, which can lead to the formation of spin-spiral ground states \cite{Bode2007, Udvardi2008, 
Menzel2012, Blugel2016} or,  in the presence of external magnetic field, to the emergence of magnetic skyrmions in chiral bulk magnets  \cite{Muhlbauer2009,Munzer2010,Yu2010}  and in  ultrathin magnetic films \cite{Yu2010a,kiselev2011,sampaio2013,romming2013}. 
Although the bilinear Heisenberg model provides a sufficient description of a large number of magnetic materials, it has been reported that higher order spin-spin interactions are necessary to account for the complex magnetic ordering in several systems \cite{Kurz2001,Heinze2011,Laszloffy2019,Hoffmann2020,Kronlein2018}.

In order to avoid using spin models the magnetic ground state of nanostructures can directly be determined from first principles calculations based on density functional theory.  The real-space linear muffin-tin orbital  method in the atomic sphere approximation (RS-LMTO-ASA) extended to non-collinear magnetic systems has been succesfully used to obtain the magnetic configurations of supported nanoclusters\cite{Bergman2006, Bergman006SS, Bergman2007}. Including the effects of spin-orbit coupling, this parameter-free method was applied to explore the complex magnetic ground state of Mn nanowires on Ag(111) and Au(111) \cite{Cardias2016}, as well as of Cr nano-islands on Pd(111) \cite{de_Melo_Rodrigues_2016}. 
Using the constrained local moment method employed with the multiple scattering Green's function technique\cite{Stocks1998,Ujfalussy1999} the existence of a 
canted magnetic state of a Co chain along a Pt(111) surface step edge \cite{Ujfalussy2004} was demonstrated from first principles in good agreement with experiment \cite{gambardella2002}.
Deriving explicit expressions for the torque acting on the magnetic moments,
the formation of a domain wall through a nano-contact has been studied within an \textit{ab-initio} framework \cite{Balogh2012}.
Using the same approach for the torque
and by solving the stochastic Landau-Lifshitz-Gilbert equations, it became possible to perform finite temperature spin-dynamics simulations for monatomic Co chains on top of Au(001) \cite{Rozsa2014}.

In this work a modified version of the conjugate gradient and Newton-Raphson methods used in \cite{Balogh2012} is proposed to find the magnetic ground state of deposited clusters from first principles.
First the numerical method will be described in details, then it will be applied to Fe chains deposited on the (111) surface of face centered (fcc) Rh. The magnetic ground states derived from the new {\em ab-initio} optimization technique are compared to those obtained from spin models parametrized from first principles and the effect of the different kind of spin-spin interactions is discussed. 
Further applications are presented for Fe chains deposited on the (110) surface of body-centered cubic (bcc) Nb and on the (0001) surface of hexagonal closed pack (hcp) Re, where spin-spiral ground states are found in agreement with previous theoretical works\cite{Laszloffy2019,Laszloffy2021} and experiment\cite{Kim2018}. 
 

\section{Methods}\label{sec:method}

\subsection{Embedded cluster calculations}

The electronic structure of the host system, composed of a semi infinite substrate, a finite number of substrate and empty sphere layers describing the surface region and a   
semi infinite vacuum, has been calculated in the terms of the fully relativistic screened Korringa-Kohn-Rostoker (SKKR) method\cite{skkr}.
Then the Green’s function embedding technique based on multiple scattering theory \cite{Lazarovits2002}
has been applied to determine the electronic and magnetic properties of the finite atomic clusters. 
 The charge and spin density have been relaxed not only for the magnetic atoms forming the chain but also in the first shell of atomic cells around them in the substrate and in the surrounding vacuum. For the construction of the effective potentials, $V_{\mathrm{eff},i}(\mathbf{r})$  and exchange-correlation fields, $\mathbf{B}_{\mathrm{xc},i}(\mathbf{r})$, where $i$ labels atomic positions, the atomic sphere approximation (ASA) has been applied. 
The exchange-correlation field in the atomic sphere corresponding to site \textit{i} is then given by 
$\mathbf{B}_{\mathrm{xc},i}(\mathbf{r}) = B_{\mathrm{xc},i}(r) \, {\bf s}_i $, where ${\bf s}_i$ is a unit vector
representing the direction of $\mathbf{B}_{\mathrm{xc},i}$.
An angular momentum cut-off $\ell_{\rm max}=3$ was used for the partial waves in multiple scattering theory. The details of the geometry used in the calculations will be specified for each system under consideration. 

\subsection{The modified conjugate gradient method}

In order to find the magnetic configuration of a cluster of atoms, a simultaneous iteration of the direction of
the magnetization, of the effective potential and of the exchange-correlation field works flawlessly in case of a simple systems. However, in case of complex magnetic structures, e.g. for frustrated antiferromagnetic or for spin-spiral ground states, the convergence of the above scheme might be less satisfactory.
An estimate of the first and second order change of the energy of the system with respect to the change of the orientation of the magnetization can help to find the magnetic ground state configuration.

In the spirit the magnetic force theorem \cite{Jansen1999}, $V_{\mathrm{eff},i}(r)$ and $B_{\mathrm{xc},i}(r)$ are kept fixed for small changes of the magnetic orientations. This implies that the number of the electrons in the finite cluster is not preserved, therefore, the variation of the energy of the system at zero temperature is replaced by the variation of the grand potential, $\Omega = E - \varepsilon_\mathrm{F}N$, where the energy $E$ of the system is approximated by the single particle (band) energy, 
\begin{equation}
  \Delta \Omega=\int\limits_{-\infty}^{\varepsilon_{\mathrm{F}}} \Delta n(\varepsilon) (\varepsilon-\varepsilon_{\rm F})\mathrm{d}\varepsilon= 
     - \int\limits_{-\infty}^{\varepsilon_{\mathrm{F}}} \Delta N(\varepsilon)\mathrm{d}\varepsilon\;,
\end{equation}
with $\varepsilon_\mathrm{F}$ and $N(\varepsilon)=\int_{-\infty}^{\varepsilon}n(\varepsilon')\mathrm{d}\varepsilon'$ being the Fermi energy and the integrated density of states (DOS), respectively. The main advantage of the above formula is that within the multiple scattering theory it can easily be calculated using Lloyd's formula \cite{Lloyd1967,Udvardi2003}.

It is straightforward to derive a formula for the first order change of the grand potential 
when the exchange correlation field in an atomic sphere at site \textit{i} is rotated around a unit 
vector $\mathbf{n}_i$ with an angle $\Delta\varphi_i$\cite{Udvardi2003}:
\begin{equation} \label{eq:E1}
\Delta \Omega^{(1)}_i =  \mathbf{T}_i \Delta{\varphi}_i \, ,
\end{equation}
where the torque $\mathbf{T}_i$ is defined as
\begin{equation} \label{eq:Ti}
 \mathbf{T}_i  = 
 \frac{1}{\pi} \int\limits_{-\infty}^{\varepsilon_\mathrm{F}} \mathrm{Im\,Tr}\left(\frac{i}{\hbar}
 \left[ \mathbf{J}, t^{-1}_i \right] \tau_{ii}\right) \mathrm{d}\varepsilon \, ,
\end{equation}
and $\Delta{\varphi}_i \equiv \mathbf{n}_i\Delta\varphi_i$,
$\mathbf{J}=\mathbf{L}+\mathbf{S}$ is the matrix of the total angular momentum operator, $t_i$ is the single-site 
scattering matrix and $\tau_{ii}$ is the site-diagonal matrix block of the scattering path operator (see \cite{Lazarovits2002}). Note that these matrices are defined in orbital momentum and spin space.

First, $V_{\mathrm{eff},i}$ and $B_{\mathrm{xc},i}$ are kept fixed and we search for the directions of the exchange-correlation field ${\bf s}_i$ at each site corresponding to the lowest grand potential. Once the restricted minimum of $\Omega$ is found, the set of ${\bf s}_i$ is kept frozen, while $V_{\mathrm{eff},i}$ and $B_{\mathrm{xc},i}$ are iterated according to the local spin-density approximation. The whole procedure is repeated until the error of 
the effective potential and the exchange correlation field as well as the torque on each site become smaller
than a predefined value.

During the optimization ${\bf s}_i$ is moving on the surface of a unit sphere. The minimum 
of the grand potential should be found on the direct product of $N$ unit spheres, where $N$ is the number of atoms included in the optimization process, thus the standard optimization methods working in Euclidean space doesn't apply. 
An extension of the conjugate gradient algorithm to Riemann 
manifolds has been used \cite{Antti2008,smith2013,Edelman98} as briefly summarized in the following. 
Let $\mathbf{t}_i^{(k)}$ be the torque vector which specifies the rotation for site \textit{i} at the \textit{k}-th iteration step. At the beginning of the iteration procedure $\mathbf{t}_i^{(0)}$ is set equal to $\mathbf{T}_i$ given by \eref{eq:Ti}.
In the first part of an iteration step we look for the minimum of the grand potential with respect to $\alpha$,
\begin{equation} \label{eq:ls}
 \Omega(\alpha) = \Omega\left\{O\left (\mathbf{n}_i^{(k)},\alpha t_i^{(k)} \right ) {\bf s}_i\right\} \; ,
\end{equation}
where $t_i^{(k)}=|\mathbf{t}_i^{(k)}|$, $\mathbf{n}_i^{(k)}=\mathbf{t}_i^{(k)}/{t}_i^{(k)}$ and $O(\mathbf{n},\varphi)$ denotes a rotation around vector $\mathbf{n}$ by an angle of $\varphi$. 
The vector $\mathbf{t}_i^{(k)}$ 
plays the same role as the search direction in a conventional conjugate gradient method. 
Once the minimum with respect to $\alpha$ has been found, we calculate the torque using \eref{eq:Ti} in the new configuration and define the effective torque for the next iteration step,
\begin{equation}\label{eq:search}
  \mathbf{t}_i^{(k+1)} = \mathbf{T}_i^{(k+1)} + \beta_i \mathbf{t}_i^{(k)} \, ,
\end{equation}
with the multipliers $\beta_i$ from the Polak-Ribi\`ere formula\cite{Polak1969},
\begin{equation}
 \beta_i = \frac{\left ( \mathbf{T}_i^{(k+1)} - \mathbf{T}_i^{(k)}\right ) \mathbf{T}_i^{(k+1)}}
 {\mathbf{T}_i^{(k+1)}\mathbf{T}_i^{(k)}} \, .
\end{equation}
The iterations are repeated until satisfactory convergence is achieved. 

\begin{figure}[htb]

  \centering \includegraphics[width=0.80\columnwidth]{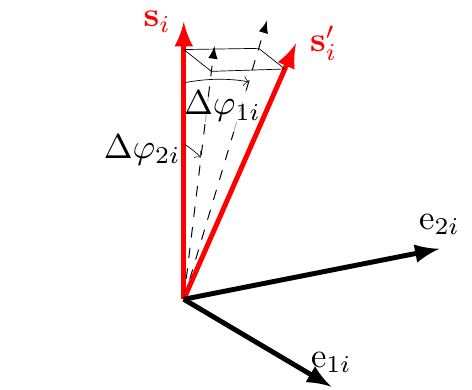}

    \caption{\label{fig:forgatas} Rotation of the magnetization vector ${\bf s}_i$ around the orthogonal directions $\mathbf{e}_{1i}$ and $\mathbf{e}_{2i}$ by angles $\Delta \varphi_{1i}$ and $\Delta \varphi_{2i}$, respectively. The new magnetization direction is denoted by ${\bf s}_i^\prime$.}
\end{figure}

In order to further increase the efficiency of the optimizing procedure, we applied a Newton-Raphson method adopted to the
Riemann manifold. 
The second order change of the grand potential necessary for the Newton-Raphson method can also be calculated 
using the Lloyd's formula \cite{Udvardi2003},
\begin{equation}\label{eq:E2}
\Delta \Omega^{(2)}_{ij}  = {\frac{1}{2}} \Delta{\varphi}_i H_{ij}\Delta{\varphi}_j \,,
\end{equation}
with the $3 \times 3$ matrix,
\begin{eqnarray}\label{eq:Hij}
H_{ij} = \frac{1}{\pi} \int\limits_{-\infty}^{\varepsilon_\mathrm{F}} \frac{1}{\hbar^2} \mathrm{Im\,Tr} 
 \left( \left[\mathbf{J}, t^{-1}_i\right]\tau_{ij}\left[\mathbf{J}, t^{-1}_j\right]\tau_{ji}\right. \nonumber \\
 \left. - \delta_{ij} \left\{ \left[\mathbf{J}, \left[\mathbf{J}, t^{-1}_i\right]\right] \right\}\tau_{ii}
  \right) \mathrm{d}\varepsilon
 \, .
\end{eqnarray}
The change of the grand potential is then given up to second order by
\begin{eqnarray}
 \Delta \Omega &= \sum_i \Delta \Omega^{(1)}_i +\sum_{ij} \Delta\Omega^{(2)}_{ij} = \nonumber\\
  &=\sum_i \mathbf{T}_i \Delta{\varphi}_i + 
 \frac{1}{2}\sum_{ij} \Delta{\varphi}_i H_{ij}\Delta{\varphi}_j\;.
 \label{eq:DeltaOmega}
\end{eqnarray}
The component of the total angular momentum operator parallel to ${\bf s}_i$ commutes with the
single site scattering matrix $t_i$, $[{\bf s}_i\mathbf{J},t_i]=0$. Consequently the following relationships apply,
\begin{eqnarray}
 \mathbf{T}_i {\bf s}_i=0, \\
  H_{ij} {\bf s}_j = {\bf 0} \, .
\end{eqnarray}
This implies that  a set of rotation $\displaystyle \left\{ \Delta {\varphi}_i \right\}$ lowering the 
grand potential 
must be constrained to the manifold perpendicular to $\displaystyle \left\{ {\bf s}_i \right\}$.
For that reason two auxiliary unit vectors per site, $\mathbf{e}_{1i}$ and $\mathbf{e}_{2i}$, are introduced  which 
together with ${\bf s}_i$ form a right-handed orthonormal system as depicted in figure \ref{fig:forgatas} and the rotations  are chosen as 
\begin{equation}
 \Delta{\varphi}_i = \Delta\varphi_{1i} \mathbf{e}_{1i} + \Delta\varphi_{2i} \mathbf{e}_{2i} \,.
\end{equation}

Varying the change of the grand potential (\ref{eq:DeltaOmega}) with respect to $\Delta\varphi_{\gamma i}$ $(\gamma=1,2)$, we obtain the Newton equations,
\begin{eqnarray}\label{eq:NR}
& T_{\gamma i} + \sum_{\gamma^\prime j}  \mathcal{H}_{\gamma i,\gamma^\prime j} \Delta\varphi_{\gamma^\prime j} \equiv \nonumber\\
& \mathbf{e}_{\gamma i}\mathbf{T}_i +
\sum_{\gamma^\prime j} \mathbf{e}_{\gamma i} H_{ij} \mathbf{e}_{\gamma^\prime j}\Delta\varphi_{\gamma^\prime j} =   0 \, .
\end{eqnarray}
Inverting the Hessian, $\mathcal{H}=\left\{ \mathcal{H}_{\gamma i,\gamma^\prime j} \right\}$, a new set of $\Delta\varphi_{\gamma i}$ is calculated, 
\begin{equation}\label{eq:newrot}
 \Delta\varphi_{\gamma i} = - \sum_{\gamma^\prime j}  \mathcal{H}^{-1}_{\gamma i,\gamma^\prime j} T_{\gamma^\prime j} \, ,
\end{equation}
thus a new spin-configuration is generated. The Newton-Raphson method converges to the local minimum
if the $\mathcal{H}$ is positive definite.
In our implementation, at the end of every line search of the conjugate gradient procedure the Hessian was checked and, if all the eigenvalues were positive, the optimization was continued using the Newton-Raphson method. 
Beyond the stable spin-moments of the iron atoms, we also included the induced moments of the non-magnetic host atoms in the minimization of the grand potential. Nevertheless, if the induced moment on a non-magnetic site was smaller than a threshold value (chosen as $0.01 \mu_{\rm B}$), the direction of the exchange field at this site was set in the next iteration parallel to the direction of the magnetization prescribed by the local spin-density approximation. 

In all cases the optimization procedure has been done until
the relative error of the effective potential and of the magnitude of the exchange-correlation field became  
smaller than $10^{-9}$ and $10^{-7}$, respectively and the overall torque on the sites,
$T = \displaystyle \sqrt{ \sum_i T_i^2}$, decreased below $10^{-2}\;\mathrm{mRyd}$. 

\subsection{Spin model calculations}

We compared the results of the first principles optimization process to those obtained from the following spin model,
\begin{equation} \label{eq:Heis}
 H= -\frac{1}{2}\sum\limits_{ij}\mathbf{s}_i \mathbf{J}_{ij}\mathbf{s}_j+ \sum\limits_i \mathbf{s}_i \mathbf{K}_i\mathbf{s}_i\;,
\end{equation}
where $\mathbf{s}_i$ and $\mathbf{s}_j$ represent the direction of the magnetic moments at site $i$ and $j$ of the magnetic cluster, $\mathbf{K}_i$ is the onsite anisotropy tensor and $\mathbf{J}_{ij}$ is a $3\times 3$ matrix, called the tensorial exchange coupling.  $\mathbf{J}_{ij}$ can be decomposed as \cite{Udvardi2003},
\begin{eqnarray}
\mathbf{J}_{ij}&=&\frac{1}{3}\textrm{Tr}\left(\mathbf{J}_{ij}\right)\mathbf{I}+\frac{1}{2}\left(\mathbf{J}_{ij}-\mathbf{J}^{T}_{ij}\right)\nonumber
\\
& \ &+\frac{1}{2}\left(\mathbf{J}_{ij}+\mathbf{J}^{T}_{ij}-\frac{2}{3}\textrm{Tr}\left(\mathbf{J}_{ij}\right)\mathbf{I}\right),\label{Jij}
\end{eqnarray}
where $J_{ij}=\frac{1}{3}\textrm{Tr}\left(\mathbf{J}_{ij}\right)$ is the isotropic Heisenberg exchange coupling, $ \mathbf{s}_{i} \frac{1}{2} \left(\mathbf{J}_{ij}-\mathbf{J}^{T}_{ij}\right)\mathbf{s}_{j}=\mathbf{D}_{ij}\left(\mathbf{s}_{i}\times\mathbf{s}_{j}\right)$ is the antisymmetric DMI, and the third, symmetric traceless part of $\mathbf{J}_{ij}$ stands for the pseudo-dipolar anisotropy. 
These parameters were obtained using the spin cluster expansion\cite{Drautz2004, Szunyogh2011, Deak2011,Laszloffy2017}, 
where the self-consistent effective potentials and exchange-correlation fields of the magnetic cluster and the neighboring host region were employed.

The ground state of the spin model was determined by consecutive Metropolis and Landau-Lifshitz-Gilbert (LLG) spin dynamics simulations \cite{Rozsa2014}. For each system ten simulations were performed from independently chosen random initial configurations. For the considered systems we found that all the ten simulations ended in the same spin-configuration (or to its time-reversed counterpart) within a relative error of $10^{-6}$. 

\section{Results}\label{sec:results}

\subsection{Fe/Rh(111)}

\begin{figure}[htb]

  \centering \includegraphics[width=0.80\columnwidth]{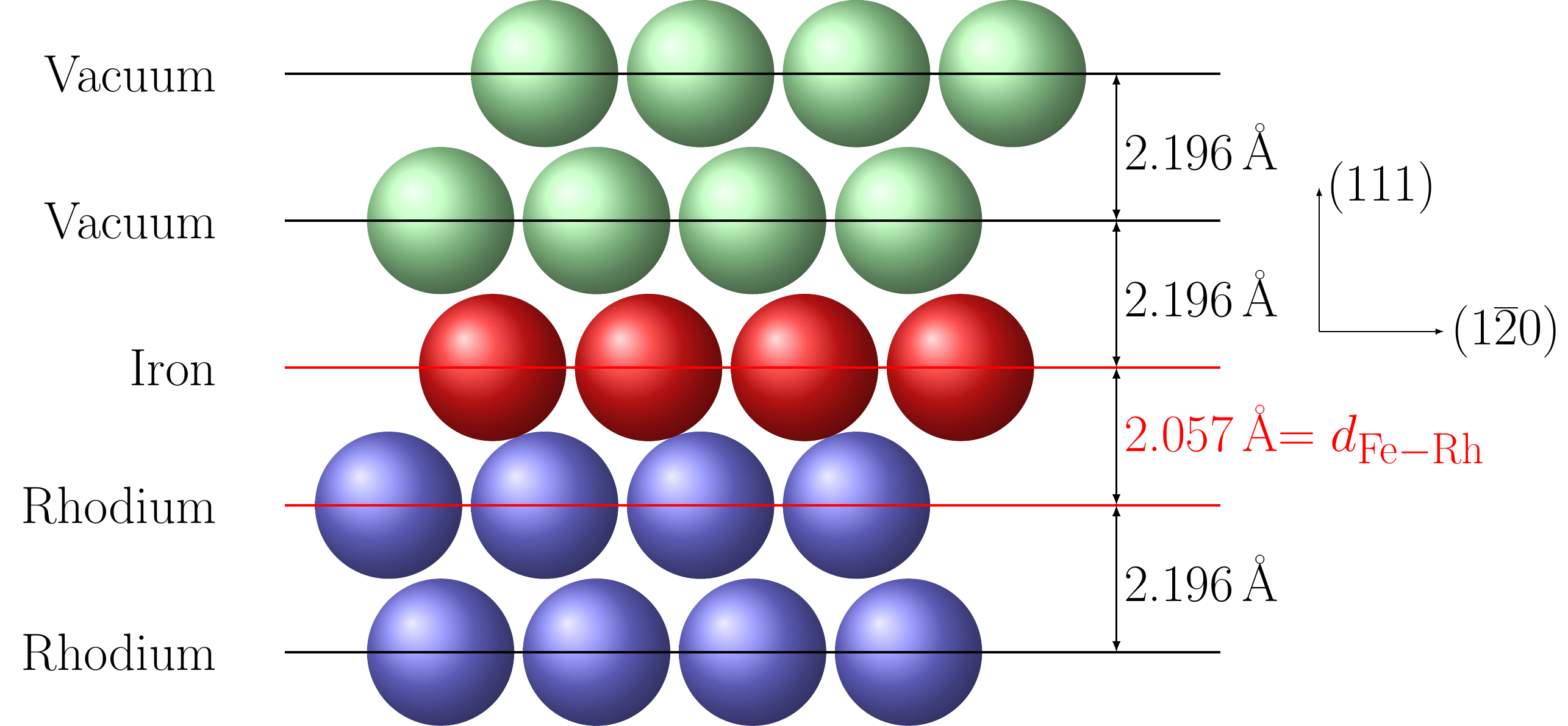}

    \caption{\label{fig:rhreteg} Sketch of the simplified layer geometry for embedded Fe chains on Rh(111) substrate. 
    The Rh and Fe atoms, as well as the empty spheres in the vacuum region are shown with blue, red and green spheres, respectively. 
    }
\end{figure}

Rh crystallizes in fcc  structure 
with  a lattice constant of $a_\mathrm{Rh}=3.803\,$\AA\, implying an inter-layer distance of 2.196\,\AA\, between the Rh(111) planes. For the vertical position of the Fe chains with respect to the uppermost Rh monolayer we used the interlayer distance between an Fe monolayer and the top Rh monolayer of the Rh(111) substrate, $d_\mathrm{Fe-Rh}=2.057\,$\AA\, (corresponding to an inward relaxation of -7.3~\%), reported by  Lehnert \emph{et al.} \cite{Lehnert2010}. The relaxation of the interlayer distances between the Rh layers found to be 1~\% or less  are neglected in our calculations for simplicity, see figure \ref{fig:rhreteg}. It should be noted that in \cite{Lehnert2010} the hexagonal close-packed (hcp) stacking position of the Fe monolayer has been found slightly lower in energy than the fcc stacking. In  \cite{blonski2010}  it was demonstrated that the magnetocrystalline anisotropy energy (MAE) is remarkably sensitive to the stacking position of Fe and Co adatoms on Rh(111) and Pd(111) surfaces. Since the MAE and, in general, relativistic effects are of great importance in determining the magnetic state of supported nanoclusters,  we investigated the magnetic ground state of monatomic Fe chains occupying both fcc and hcp stacking positions above the topmost Rh monolayer. 

The electronic structure of the host Rh(111) surface was calculated using the SKKR method\cite{skkr}.
The Rh(111) surface was modeled by eight monolayers of Rh and four monolayers of empty spheres (vacuum) between semi-infinite bulk Rh and semi-infinite vacuum regions. The vertical geometry of the layers was identical with that shown in figure \ref{fig:rhreteg}, the Fe monolayer being replaced by a layer of empty spheres.    
The iron chains were then embedded into this vacuum layer by means of the embedding technique based on the KKR Green's function method\cite{Lazarovits2002} and self-consistent calculations of the magnetic ground states have been performed as described in Section II.
We studied close-packed chains of 4, 5, 6 and 7 Fe atoms deposited in the $[1\overline{1}0]$ direction, where the Fe atoms occupied either the fcc or the hcp hollow positions above the Rh(111) surface.

\begin{figure*}[htb]  
  \centering \includegraphics[width=1.99\columnwidth]{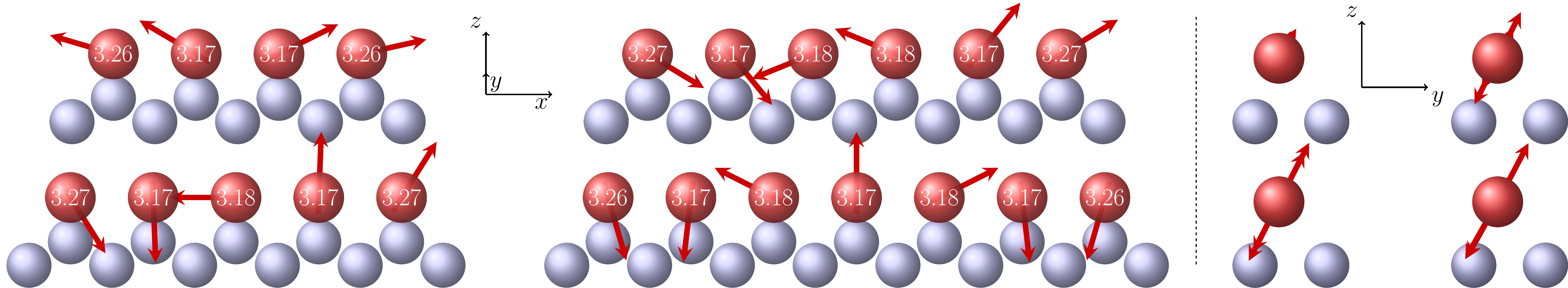}      
  \caption{\label{fig:fcc_egybe_4-7} Magnetic ground states of monatomic chains of 4, 5, 6 and 7 Fe atoms on a Rh(111) surface with fcc stacking obtained from \textit{ab-initio} optimization. 
  The $x$, $y$ and $z$ axes correspond to the $[1\overline{1}0]$, $[1\overline{1}\overline{2}]$ and $[111]$ directions, respectively Red spheres represent the Fe atoms, while the blue ones represent atoms in the underlying Rh layer. The red vectors show the direction of the spin magnetic moments of the Fe atoms, while their magnitude in $\mu_\mathrm{B}$ is written inside the red spheres. The panels on the right show the 
  normal-to-chain views of the spin-con\-fi\-gu\-ra\-tions of the four considered chains in the same order as in the left panels.
      }
\end {figure*}
In figure \ref{fig:fcc_egybe_4-7} we present the ground state magnetic configuration for the considered Fe chains with fcc stacking. The magnitudes of the Fe magnetic moments inside the chain show slight oscillations in the range of 3.17-3.18 $\mu_\mathrm{B}$, while the edge atoms exhibit a modestly enhanced moment of 3.26-3.27 $\mu_\mathrm{B}$ which can be attributed to the reduced coordination of the edge Fe atoms.
The magnitude of the induced magnetic moments of the Rh atoms nearest the Fe atoms (not displayed in figure \ref{fig:fcc_egybe_4-7}) is around  $0.2\,\mu_\mathrm{B}$.

Before discussing the ground-state spin-con\-fi\-gu\-ra\-tions displayed in figure \ref{fig:fcc_egybe_4-7} we note that the point group of the chains contains two operations: the identity and a mirror plane perpendicular to the chain. In addition, time reversal that turns around the spin vectors is also a symmetry operation of the system. Accordingly, there exist two kinds of ground-state spin-configurations being invariant either under the mirror transformation or under simultaneous action of mirror transformation and time reversal. It then follows that the spin vectors can transform either as axial or polar vectors under the mirror plane operation.

As can be seen from figure \ref{fig:fcc_egybe_4-7}a, either of the above symmetry of the magnetic configurations can be obtained from the {\it ab-initio} optimization depending on the length of the chains: the 4- and 7-atom-long chains show polar vector symmetry, while the magnetic moments in the 5- and 6-atom-long chains  behave as axial vectors. Inferring the ground-state spin-configurations, it is tempting that the building blocks of these configurations are pairs of Fe spins being aligned closely parallel to each other ($\uparrow \uparrow$), while these dimers seem to be coupled antiparallel to each other ($\uparrow \uparrow \downarrow \downarrow$). This structure is particularly obvious for the chains containing even number of Fe atoms, as what follows referred to as even chains.  In case of chains containing odd number of Fe atoms (odd chains) the alternation of  $\uparrow \uparrow$ and $\downarrow \downarrow$ dimers is necessarily broken, which gives rise to strongly  noncollinear spin-configurations. This might be associated with the fact that the central spin in an odd chain must be aligned either normal to the chain (polar vector symmetry) or parallel to the chain (axial vector symmetry).  Apparently, noncollinearity of the spins is present in the even chains too. Furthermore, all the ground state spin-configurations 
are coplanar with the planes tilted to the $z$ axis by roughly the same angle.

As what follows we make an attempt to analyze the main features of the ground-state spin configurations for the different chains in terms of the bilinear spin model \eref{eq:Heis}.
The leading terms in the spin model are the isotropic exchange interactions.  In Table \ref{tbl:jij-k} we present the nearest neighbor (NN) and next nearest neighbor (NNN) isotropic interactions between the Fe atoms. Apparently, the NN and NNN isotropic interactions are ferromagnetic (FM) and antiferromagnetic (AFM), respectively. 
These interactions are consistent with the alternating $\uparrow \uparrow\downarrow \downarrow$ structure seen as the dominant feature of the even chains. For $n=6$ the large FM coupling between the inner atoms, $J_{34}$, further stabilizes this spin structure. In the case of odd number of atoms this double-pair-wise antiferromagnetic configuration is obviously broken. 
It should be noted that a $\uparrow \uparrow\downarrow \downarrow$ double-row-wise AFM structure has been theoretically predicted \cite{hardrat2009-FeRh111} and experimentally observed \cite{kroenlein2018-FeRh111} in an Fe monolayer on Rh(111). While the emergence of such a spin-structure in a monolayer requires the presence of higher-order (in particular, three-site four-spin) interactions \cite{kroenlein2018-FeRh111}, for a chain of finite length it can be stabilized by the inhomogeneous bilinear isotropic interactions only.

\begin{table}[htb!]
 \caption{\label{tbl:jij-k} Nearest neighbor and next nearest neighbor isotropic exchange interactions $J_{ij}$ (in mRy) between sites $i$  and $j$ in fcc-stacked monatomic chains of $n$ Fe atoms. Label 1 denotes atoms at the left edge of the chain. Only the independent interactions are shown.
 }
\begin{indented}
 \item[]
\begin{tabular}{ccccc}
\br
   $i-j$&   $n=4$    & $n=5$   &  $n=6$   &  $n=7$ \\ \mr
   $1-2$&\phantom{-}0.604   & \phantom{-}0.572&\phantom{-}0.686 & \phantom{-}0.605 \\
   $1-3$&   -0.562          & -0.624          & -0.614           &-0.599\\ 
   $2-3$&\phantom{-}0.114   & \phantom{-}0.331&\phantom{-}0.558 & \phantom{-}0.322\\ 
   $2-4$&                   & -0.441          & -0.456          &-0.406\\ 
   $3-4$&                   &                 &\phantom{-}1.181 & \phantom{-}0.227\\ 
   $3-5$&                   &                 &        &-0.420\\ 
   \br
 \end{tabular}
 \end{indented}
\end{table}

\begin{figure*}[htb!]
  \centering 
  \includegraphics[width=0.860\columnwidth]{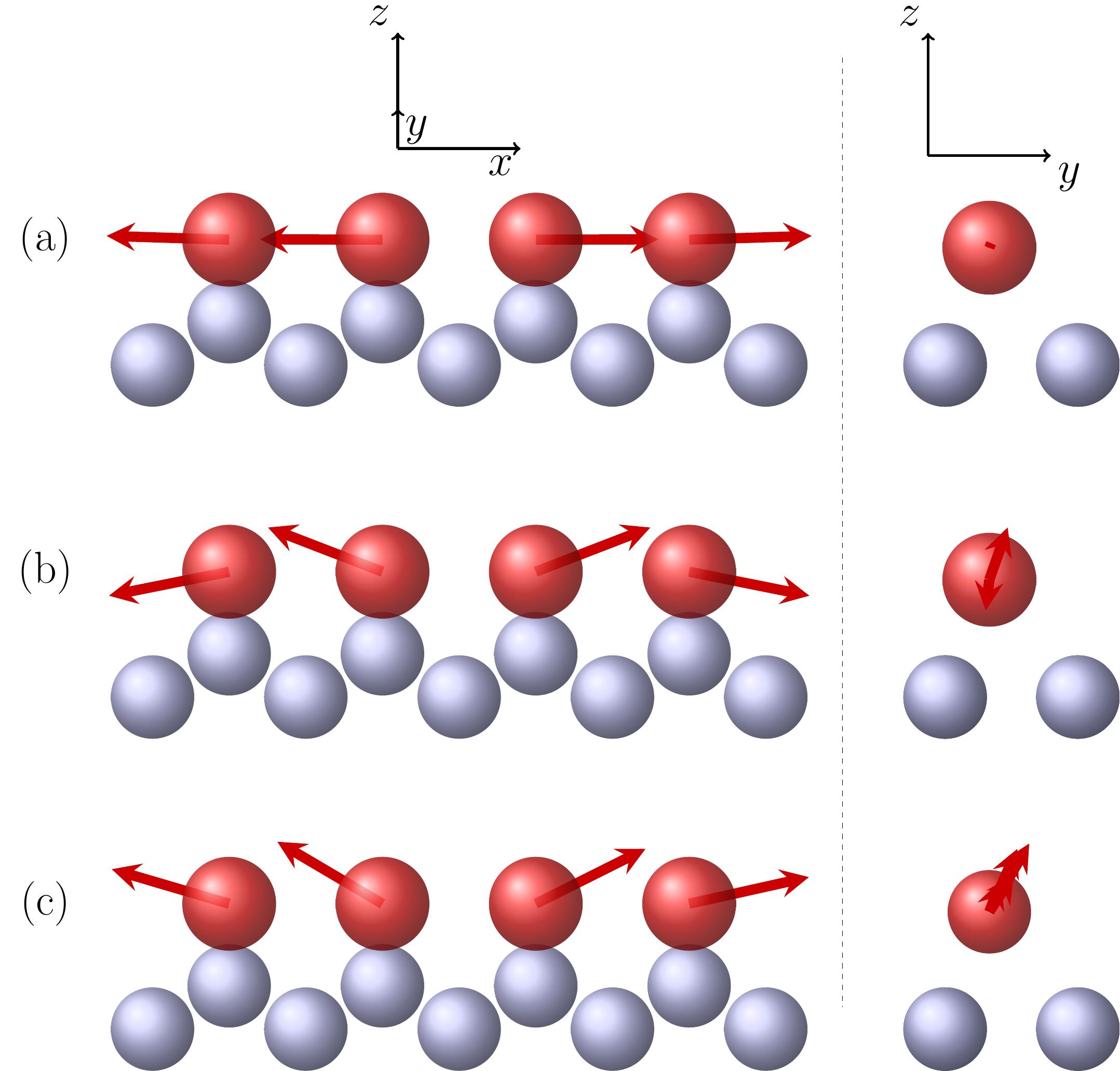}
  \hskip  1cm
   \includegraphics[width=0.980\columnwidth]{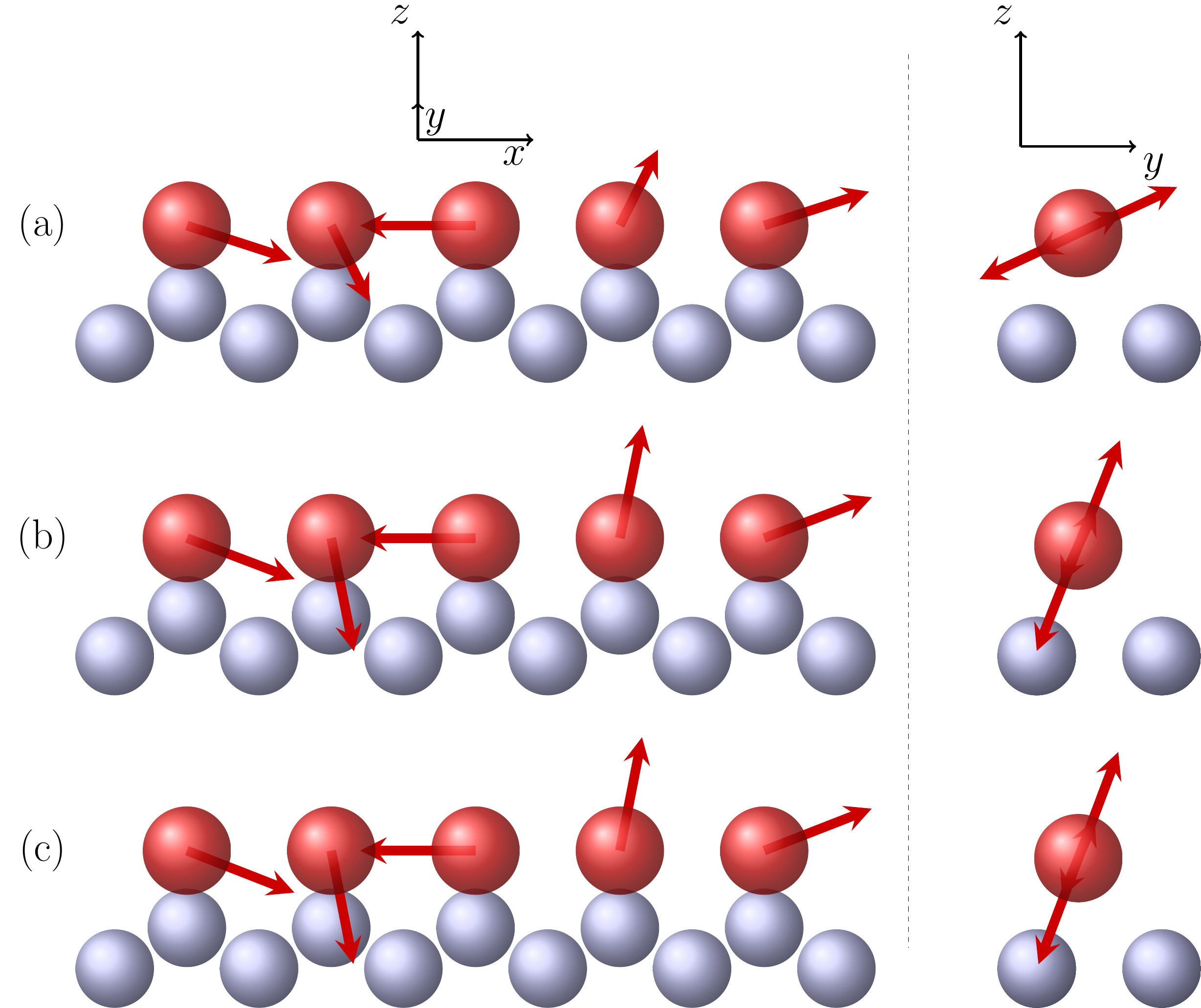}    
      \caption{\label{fig:spinmodel_lanc4-5} Ground-state spin configurations of a 4-atom-long Fe chain (left) and a 5-atom-long Fe chain (right) in fcc-stacked positions on  Rh(111) surface based on spin models as follows: 
 (a) isotropic exchange interactions and on-site anisotropy, (b) the previous spin model extended with DMI, and (c) full tensorial interactions. 
}
 \end {figure*}

For the cases of the 4-atom-long and 5-atom-long chains we demonstrate the effect of different kinds of exchange interactions in forming the magnetic ground state. For this reason, we consecutively considered  three sets of spin-model parameters: (a) isotropic interactions and on-site anisotropy matrices, (b) the previous spin-model extended by DMI and (c) the spin-model with full tensorial interactions  \eref{eq:Heis}.  
The spin-configurations obtained by spin-dynamic simulations using these spin-models are presented in 
figure~\ref{fig:spinmodel_lanc4-5}.

In case of the 4-atom-long chain, see figure~\ref{fig:spinmodel_lanc4-5} left panel, when only isotropic exchange interactions and on-site anisotropies are present, the simulations resulted in to a nearly collinear 
double-pair-wise AFM configuration along the chain direction due to an easy $x$-axis anisotropy. This configuration clearly respects a polar-vector symmetry of the spins. The $y$-components of the DM vectors introduce non-collinearity of the spin-structure in the $x$-$z$ plane by keeping the polar-vector symmetry. In addition, the plane of the spins is slightly rotated to the $y$ axis, which shows the preference of the $y$ axis with respect to the $z$ axis by on-site anisotropy  (for a discussion of the tilting of spin-spiral states see \cite{Laszloffy2019}). Switching on the symmetric exchange anisotropy doesn't considerably affect the spin-configuration.

As expected, in case of the 5-atom-long chain, see figure~\ref{fig:spinmodel_lanc4-5} right panel, the frustration of the isotropic exchange interactions causes a strongly non-collinear spin arrangement, with the middle spin oriented along the chain, again due to easy $x$-axis anisotropy. The ground state exhibits, therefore, an axial-vector symmetry. 
The $y-z$ view of the spin configuration shows that the plane of the spins is slightly tilted away from the $x-y$ plane, which is mainly due to the nonvanishing $xz$ components of the on-site anisotropy matrices. 
Switching on the DMI further stabilizes this configuration by just slightly changing the relative angles between the spins. On the other hand, the plane of the spins is tilted close to the $x-z$ plane which can be attributed to large $y$ components of the DM vectors \cite{Laszloffy2019}.
As for the 4-atom-long chain, including the symmetric exchange anisotropy has negligible effect on the magnetic ground state.

\begin{table}[htb!]
 \caption{\label{tbl:cgrad-model-dev} The mean deviation $\Delta$, defined by \eref{eq:Delta} between the magnetic ground states obtained from \textit{ab initio} optimization and from the spin model \eref{eq:Heis}.  }
\begin{indented}
 \item[]
 \begin{tabular}{cccc}
 \br
   $n=4$ & $n=5$ & $n=6$  & $n=7$ \\ \mr
   0.332 & 0.417 &  0.297 & 0.601 \\
   \br
 \end{tabular}
\end{indented}
\end{table}

The ground state configurations of the Fe chains on Rh(111) obtained from \textit{ab initio} optimization and based on the bilinear tensorial spin model compare remarkably well. However, in particular, for the longer chains larger differences occur mostly at the ends of the chains.
In order to quantify the differences between the magnetic configurations obtained from these two methods we use the average mean deviation of the spin vectors $\Delta$, the square of which is defined as
\begin{equation}
        \Delta^2=\frac{1}{n}{\sum_{i=1}^n(\mathbf{s}^\mathrm{m}_i-\mathbf{s}^\mathrm{a}_i)^2} \;,
        \label{eq:Delta}
\end{equation}
with $\mathbf{s}^\mathrm{m}_i$ and $\mathbf{s}^\mathrm{a}_i$ being the spin vectors at site $i$ obtained from the spin model and the \textit{ab initio} optimization, respectively.
The mean deviations for the different chains are summarized in Table~\ref{tbl:cgrad-model-dev}. It is apparent that the agreement between the magnetic ground states based on {\em ab initio} optimization and on the spin model is significantly better for even chains which might be attributed to the missing geometrical frustration, thus to reduced non-collinearity in these systems.

\begin{figure}[htb!]
  \centering \includegraphics[width=0.99\columnwidth]{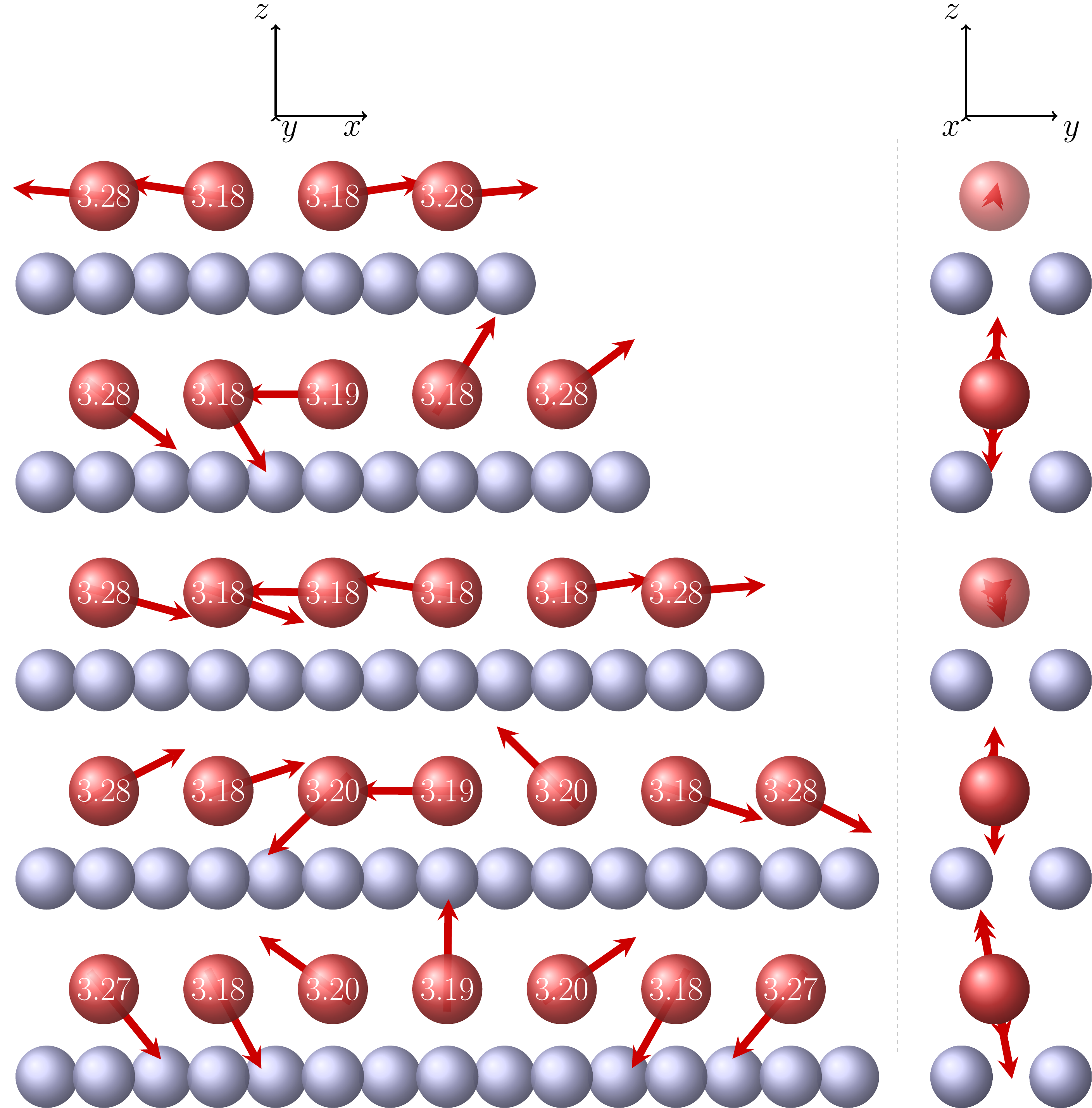}   
  \caption{\label{fig:hcp_egybe_4-7} Magnetic ground state of monatomic chains of 4, 5, 6 and 7 Fe atoms on Rh(111) surface with hcp stacking obtained from \textit{ab-initio} optimization. 
  For the details of the figure see the caption of figure~\ref{fig:fcc_egybe_4-7}.  The fourth entry from the top shows the magnetic ground state of the 7-atom-long chain with axial vector symmetry, while the last entry depicts a metastable state of this chain with polar vector symmetry. 
 
      }
\end {figure}

As we mentioned before we also applied the {\em ab initio} optimization technique to find the magnetic ground state of the Fe chains in hcp stacking. Our results are shown in figure \ref{fig:hcp_egybe_4-7}.  The magnitudes of the spin moments are very similar to those obtained for the chains with fcc stacking. Concerning the ground-state spin configurations, the double pair-wise AFM structure keep dominating, which indicates that the isotropic interactions are similar for both kinds of stacking geometry of the Fe chains.  Seemingly, the even chains remain almost collinear. This behavior could be attributed to relatively weak DM interactions compared to strong isotropic interactions. However, a spin-model study as in case of the fcc-stacked chains yields a ground-state configuration for the 6-atom with enhanced non-collinearity, implying that the bilinear spin model \eref{eq:Heis} is less suitable to recover the magnetic ground state obtained from the {\it ab initio} optimization. 

As in case of fcc stacking the odd chains with hcp stacking show a largely non-collinear magnetic ground-state, while the plane of the magnetic moments is almost normal to the surface. For the 7-atom-long chain the magnetic ground state shows axial-vector symmetry, see the fourth entry in figure \ref{fig:hcp_egybe_4-7}. In addition, we also found a metastable state with polar-vector symmetry depicted in the last entry of figure \ref{fig:hcp_egybe_4-7}, which is by $16.3\,\mathrm{meV}$ higher in energy than the ground state. These two configurations are related to each other by about 90$^\circ$ global rotation around the normal of the plane of the spins, while the metastable configuration is slightly tilted from the $z$ axis. Note that both configurations have right-handed chirality due to positive $y$ components of the DM vectors.

\subsection{Fe chains on Nb(110) and Re(0001)}

\begin{figure}[htb!]
\centering
   \includegraphics[width=0.8\columnwidth]{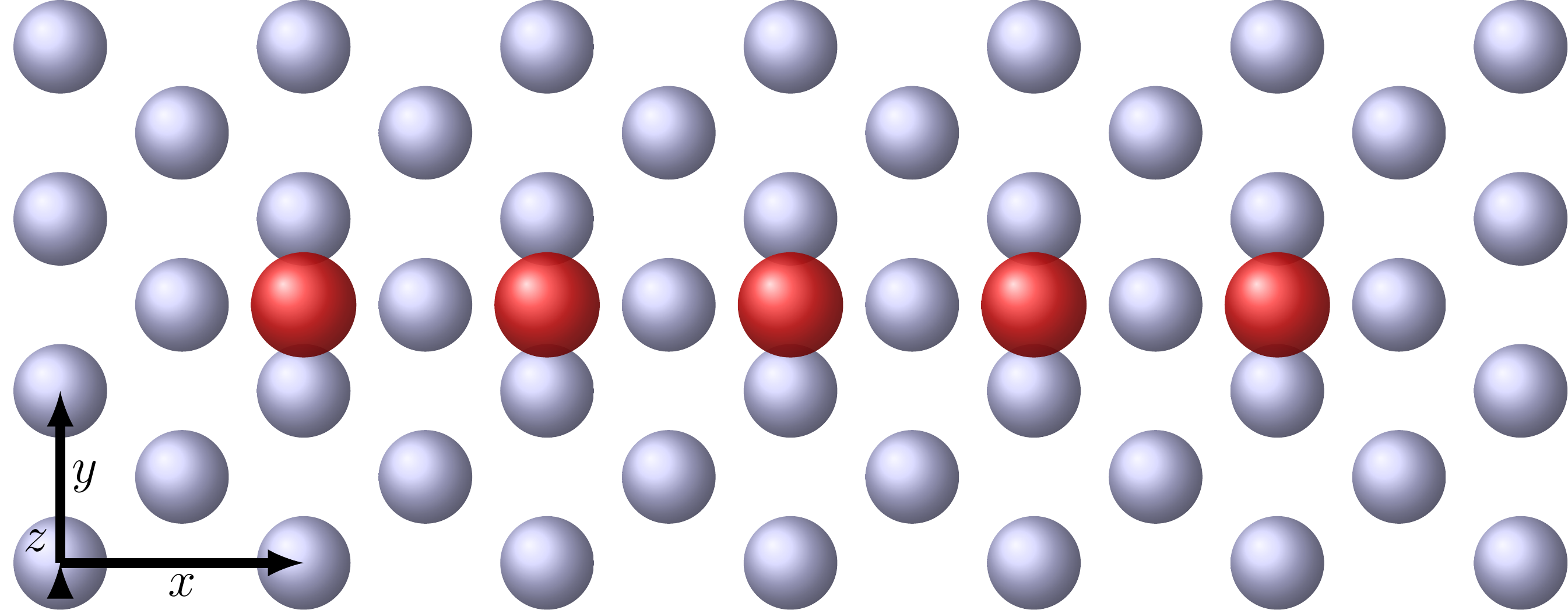}
    \caption{\label{fig:Nb_substrate}
    Top view of an array of Fe atoms on top of a Nb(110) surface studied in the present work. The Fe and Nb atoms are marked by red circles, the blue circles stand for the Nb atoms. The $x$, $y$ and $z$ axes correspond to the $[1\overline{1}0]$, [001] and $[110]$ directions, respectively. The chains are deposited along the $[1\overline{1}0]$ direction.}
\end{figure}

In this section we present our main results on the magnetic structure of Fe chains on top of Nb(110) and Re(0001) surfaces. These systems are of particular interest, since they are supposed to host Majorana edge states\cite{Kim2018,beck2021,crawford2021}.  
Bulk Nb has a bcc structure with lattice constant of $a_\mathrm{Nb}=3.3004$\,\AA.
Using the Vienna Ab-initio Simulation Package (VASP)\cite{vasp} Laszloffy {\em et al.}\cite{Laszloffy2021} found that by putting an Fe adatom on the (110) surface of Nb, the average vertical distance of the atoms in the two uppermost Nb  layers decreased to 
$2.2542$\,\AA\, from the bulk value of $2.3337$\,\AA\,, while the vertical distance between the Fe adatom and the closest Nb atoms decreased to 
$1.9868$\,\AA . In the calculations of atomic clusters of Fe  these values were used for the interlayer distances between the uppermost two Nb layers, as well as between the top Nb layer and the first two vacuum layer, in which the Fe atoms were embedded\cite{Laszloffy2021}, respectively. We adopted this geometry in our present calculations of the Fe chains in terms of the embedding Green's function technique. The host surface was modelled by eight Nb layers and four empty sphere layers sandwiched between semi-infinite bulk Nb and vacuum regions. The top view of an Fe chain on the Nb(110) surface is shown in figure \ref{fig:Nb_substrate}. 

\begin{figure*}[htb!]
   \includegraphics[width=1.99\columnwidth]{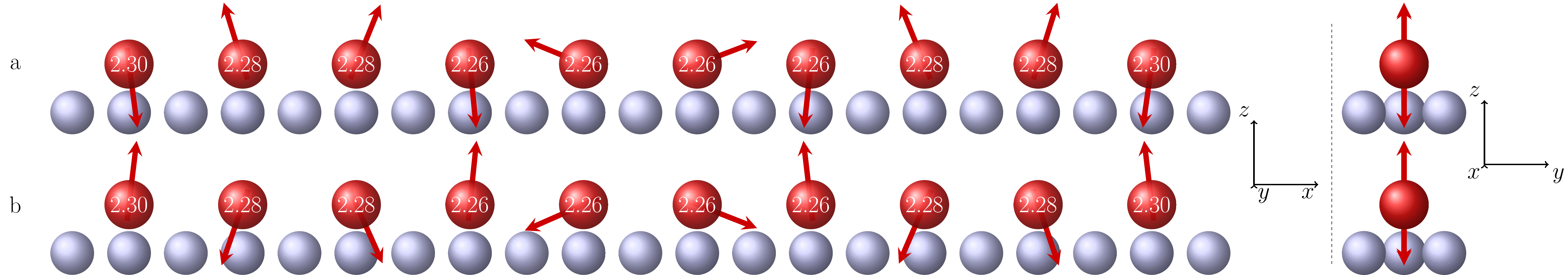}
    \caption{\label{fig:FeNb_chains}
     (a) Metastable and (b) ground state magnetic configurations obtained by {\em ab initio} optimization for a 10-atom long Fe chain on Nb(110). Red and blue circles represent the Fe and Nb atoms, respectively. 
     The red (unit)vectors show the direction of the magnetization vector, while its magnitude is written inside the red spheres in $\mu_\mathrm{B}$ units. The $x$, $y$ and $z$ directions correspond in order to the $[1\overline{1}0]$, [001] and $[110]$ directions. The pictures on the right show the $y$-$z$ side-view of the spin vectors. Note that the rotational sense of spin-spiral (a) is right-handed (clockwise) around the positive $y$ axis, while spin-spiral (b) has a left-handed (anti-clockwise) rotational direction.
  }
\end{figure*}

We optimized the magnetic ground state of 5-, 10- and 15-atom long Fe chains deposited along the $[1\overline{1}0]$ ($x$) direction. Similar to the spin-model study in  \cite{Laszloffy2021}, we found that the ground state of the investigated chains is a spin-spiral. In figure \ref{fig:FeNb_chains} we present two optimized magnetic configurations for the 10-atom long chain. Remarkably, the spin-moments of the Fe atoms is by about 1\,$\mu_{\textrm{B}}$ less than in the chains on the Rh(111) surface which is a typical volume effect. Due to the fact that the hard axis of the system is $y$, the spins rotate in the $x$-$z$ plane. 
Based on a spin model with exchange interactions obtained from first principles, in  \cite{Laszloffy2021} a spin spiral wavelength of $\lambda$=3.39\,$a_x$ was found, where $a_x=\sqrt{2}\,a_\mathrm{Nb}$ is the distance between the subsequent iron atoms in the chain. For the 10-atom-long chain, we performed a least square fit of the rotation angles of the spins with respect to $x$ axis to a linear function and we obtained a slope of 2.17 rad/site which translates into a spin-spiral wavelength of $\lambda$=2.90\,$a_x$, which means a decrease by about 15\% as compared to \cite{Laszloffy2021}.

It is obvious that the metastable state shown in figure \ref{fig:FeNb_chains}a  and the ground state shown in figure \ref{fig:FeNb_chains}b  have opposite rotational sense. The ground state is lower in energy by $1.7\,\mathrm{meV}$ than the metastable state, converted into about $0.08\,\mathrm{meV}$/Fe, which is indeed in the energy range of the DMI of these chains\cite{Laszloffy2021}. 
It is pointed out in \cite{Laszloffy2021}  that the frustrated isotropic interactions and the relatively strong anisotropy can stabilize spin spirals with opposite chirality for longer chains. It should be noted, however, that the ground state spin-spiral in figure \ref{fig:FeNb_chains}b has a left-handed (anti-clockwise) rotational sense as opposed to that obtained from the spin model\cite{Laszloffy2021} which has a right-handed (clockwise) rotational sense. This discrepancy might be attributed to higher-order chiral spin-spin interactions\cite{Laszloffy2019,Brinker2019} missing in the spin-model used in \cite{Laszloffy2021}.

Finally, we turn to Fe chains deposited on top of the (0001) surface of hcp Rhenium. In our calculations we used the bulk in-plane lattice constant of $a_\mathrm{2D}=2.761$\,\AA \ and the interlayer relaxations found by Lászlóffy \textit{et al.} \cite{Laszloffy2019} in terms of the VASP code\cite{vasp}.
As compared to the interlayer distance $2.281$\,\AA \ of the (0001) planes in bulk Re, the distance between the uppermost Re layer and the Fe adatoms decreased to $2.228\,$\AA, while the distance between the surface and subsurface Re layers reduced to $2.16\,$\AA.
The Re(0001) surface was modeled as an interface region between a semi-infinite bulk Re and vacuum consisting of 8 monolayers of Re and four monolayers of empty spheres (vacuum). The iron chains were embedded into the lowermost vacuum layer along the nearest neighbor direction ($x$). 

\begin{figure*}[htb!]
   \includegraphics[width=1.99\columnwidth]{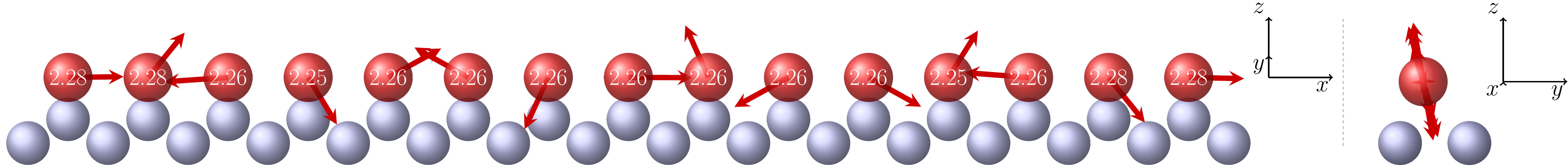}
    \caption{\label{fig:re_spiral} 
     Ground-state spin-configuration of a 15-atom long Fe chain on Re(0001) substrate.  The red spheres correspond to the Fe atoms, while the blue ones belong the underlying Re layer. The $x$, $y$ and $z$ directions correspond to the $[1000]$, $[1\overline{2}00]$ and $[0001]$ directions of the hcp crystal structure, respectively. Red vectors show the directions of the atomic magnetic moments, while their magnitudes are written inside the spheres in $\mu_\mathrm{B}$ units.
      }
\end{figure*}

The ground state magnetic configuration of a 15-atom-long chain of Fe atoms obtained via {\em ab initio} optimization is shown in figure \ref{fig:re_spiral}. This ground state is obviously a similar spin-spiral state as found by Laszloffy {\em et al.}\cite{Laszloffy2019} also by {\em ab initio} spin-dynamics using a cut-off $\ell_{\mathrm{max}}=2$ for the multiple scattering calculations (see figure~5(e) of \cite{Laszloffy2019}). 
A linear fit of the rotational angles of the spin vectors in figure \ref{fig:re_spiral} yields a spin-spiral wavelength of about $3.4\,a_\mathrm{2D}$.  Calculations on chains of 11, 13, 17 and 19 Fe atoms resulted in similar wavelengths. This wavelength is considerably lower than the value of about $5\,a_\mathrm{2D}$ obtained in \cite{Laszloffy2019}. Nevertheless,
both values are close to $\lambda = 4\,a_\mathrm{2D}$ found experimentally using spin polarized scanning tunneling microscopy by Kim {\em et al.}\cite{Kim2018} for a 40-atom long chain.

\section{Summary}
We performed fully converged unconstrained self-consistent calculations of the magnetic ground state of  Fe chains deposited on different metallic surfaces by using a newly develepod method based on a combination of the conjugate gradient and Newton-Raphson methods within the multiple scattering Green's function embedding technique.
In case of Fe chains in fcc stacking positions on top of Rh(111) we found that the building block of the obtained magnetic configurations is a double pair-wise antiferromagnetic structure that subsists for chains of even number of Fe atoms, while for odd number of Fe atoms the geometrical frustration leads to strongly non-collinear magnetic ground states. We established that, as the length of the chain is varied, spin-configurations with either axial or polar vector symmetry occur. Simulations based on a tensorial spin-model showed that the isotropic exchange couplings  and the onsite anisotropy sufficiently explain the main features of the magnetic ground state, but the Dzyaloshinky-Moriya interactions play an important role in the formation of the final, complex spin-configuration. A numerical  comparison of the ground state spin-configurations from the two methods showed larger deviations for higher degree of complexity (non-collinearity) of the magnetic states indicating that higher order spin-spin interactions should be included in spin-model studies. Our calculations revealed that the Fe chains in hcp stacking positions have similar spin-structures as in fcc stacking.  For the 7-atom long chain the magnetic ground state had axial vector symmetry, but we also found a metastable state with polar vector symmetry stabilized by DM interactions.

In case of Fe chains deposited on Nb(110) surface along the second nearest neighbor direction, we found almost degenerate spin-spiral states with opposite chiralities in line with a recent spin-model study by Laszloffy {\em et al.}\cite{Laszloffy2021}, who demonstrated that the quasi-degeneracy of these spin-spiral states is the consequence of the frustration of isotropic interactions and the DM interactions, while spin-spirals not preferred by the DMI are stabilized by strong magnetic anisotropy. For Fe chains on Re(0001) surface, our first principles optimization process led to spin spiral states with a wavelength comparable with the experimental finding\cite{Kim2018}.

In conclusion, we demonstrated that the optimization method introduced in this work is an efficient tool to explore the complex magnetic ground states of supported atomic clusters. This might be particularly helpful for a material specific design of new magnetic nanostructures for quantum information technology based on low-dimensional topological superconductivity.   

 
 \ack
The authors are thank András Lászlóffy for the help in the spin model calculations and Levente Rózsa for fruitful discussions.
This research was funded by the Ministry of Innovation and  the National Research, Development, and Innovation Office of Hungary under the  project No. K131938, project No. FK124100  and under Grant TKP2021-NVA-02.

\section*{References}

\bibliographystyle{iopart-num}
\bibliography{paper_clu_opt}

\end{document}